\pdfoutput=1
\documentclass[11pt]{article}
\usepackage[margin=1.1in]{geometry}
\usepackage{amsmath,amssymb,amsthm}
\usepackage{graphicx}
\usepackage{booktabs}
\usepackage{multirow}
\usepackage[numbers,sort&compress]{natbib}
\usepackage[colorlinks=true,linkcolor=blue,citecolor=blue,urlcolor=blue]{hyperref}
\usepackage{xcolor}
\usepackage{tikz}
\usetikzlibrary{arrows,positioning}
\usepackage{authblk}

\newtheorem{theorem}{Theorem}

\newtheorem{proposition}{Proposition}

\theoremstyle{remark}

\newcommand{\pr}{\mathbb{P}}
\newcommand{\ex}{\mathbb{E}}
\newcommand{\Msf}{M}

\title{ARM: Detector-Agnostic Changepoint Attribution with
Finite-Sample Error Control}

\author[1,2]{Chenchen Peng}
\author[1]{Mixia Wu}
\author[1]{Qijing Yan}
\author[3,2]{Da Chen\thanks{Corresponding author. Email:
\texttt{dachensmu@163.com}}}
\author[2]{Zhiqi Shen}

\affil[1]{School of Mathematics, Statistics and Mechanics, Beijing
University of Technology, Beijing 100124, China}
\affil[2]{College of Computing and Data Science, Nanyang Technological
University, Singapore 639798, Singapore}
\affil[3]{College of Electrical Engineering, Sichuan University, Chengdu,
Sichuan 610065, China}

\date{August 2, 2026}

\begin{document}
\maketitle

\begin{abstract}
Detecting a change in a multivariate series answers only the first of
two questions; the operational question is which coordinates changed.
Existing answers are incomplete. Block-level procedures certify
predefined groups of coordinates under an additive union bound,
high-dimensional variable-selection methods return interpretable
rankings without error guarantees, and the post-detection inference
literature controls error along the time axis rather than across
coordinates. We propose ARM (Attribution by Rank Maxima), a wrapper
that accepts a changepoint located by an arbitrary detector and returns
the set of coordinates certified to have changed, each carrying a
location or scale type label. ARM scores each coordinate by a
max-over-splits rank statistic. Because this statistic dominates the
corresponding statistic at the estimated split, the resulting
certificate is invariant to the manner, and to the accuracy, of the
changepoint estimate. Three finite-sample guarantees follow from
within-coordinate ranks alone: per-coordinate validity under any
detector; exact family-wise error control through a Westfall--Young
joint permutation that preserves cross-coordinate dependence, with a
fully distribution-free Holm fallback; and false discovery rate control
under arbitrary coordinate dependence in high dimensions through
Benjamini--Yekutieli and e-BH. In simulations, naive per-coordinate
testing at the estimated changepoint inflates its family-wise error
beyond $0.66$ as the dimension grows, whereas ARM maintains the nominal
level while retaining validity under heavy tails, power in high
dimensions, and accurate type labels. On five financial series
surrounding the 2008 collapse, ARM attributes a scale change to every
asset class and excludes injected control coordinates.
\end{abstract}

\noindent\textbf{Keywords:} changepoint attribution; post-detection
inference; rank statistics; family-wise error; false discovery rate;
permutation test

\section{Introduction}\label{sec:intro}

When a monitoring system flags a structural change in a multivariate
stream, whether a sensor array, a portfolio, or a network of meters, the
question of when the change occurred is answered by the detector itself
and refined by a substantial literature \citep{li2026modern}. The
operational question that follows is which coordinates are responsible:
which sensors drifted, which assets drove the regime shift, which
meters failed. Standard practice takes the detector's estimated
changepoint $\hat\tau$ and applies a two-sample test to each coordinate
across it. This practice is invalid. The estimate $\hat\tau$ is
computed from the same data, at the location that maximizes aggregate
evidence, so testing each coordinate at $\hat\tau$ refers a selected
statistic to an unselected null distribution and inflates the type-I
error, increasingly severely as the dimension grows;
Section~\ref{sec:sim-killshot} quantifies the inflation.

Three lines of work address related problems. The post-detection
inference literature controls error rigorously along the time axis,
testing which alarms correspond to genuine changes
\citep{jia2024tune,cui2026art} or constructing confidence statements
for the location
\citep{jewell2022post,saha2026post,saha2026distributionfree}, but does
not address the coordinate axis. For the attribution question itself,
\citet{ouerfelli2026posthoc} give the first formal treatment: kernel
two-sample tests certify whether the change occurs in either of two
predefined blocks of coordinates, with the family-wise error rate
bounded additively as $\mathrm{FWER}\le\alpha_0+L\alpha_1$ over $L$
blocks and with calibration based on a holdout split. Finally, a strand
of high-dimensional methods, exemplified by WAVE \citep{lan2026wave},
produces coordinate-level attribution by fusing adaptive weights and
local effect sizes into a ranking; a ranking is informative but carries
no finite-sample error guarantee.

What is missing is attribution formulated as a detector-agnostic
wrapper with coordinate-level finite-sample error control that remains
valid at an estimated changepoint, providing exact family-wise error
control without additive union-bound slack, false discovery rate
control robust to coordinate dependence in high dimensions, and a
change-type label for every certified coordinate. The present paper
supplies this formulation.

We propose ARM (Attribution by Rank Maxima). The method scores
coordinate $j$ by the max-over-splits rank statistic
$\Msf_j=\max_t\max\{Z^{\mathrm{loc}}_j(t),Z^{\mathrm{sc}}_j(t)\}$,
which combines Wilcoxon location and Mood scale evidence over a grid of
candidate splits. Since $Z_j(\hat\tau)\le\Msf_j$ for every split, the
certificate neither uses nor can be invalidated by the detector's
estimate (Theorem~\ref{thm:pivotal}). Three guarantees then follow from
within-coordinate ranks alone: per-coordinate validity under any
detector (Theorem~\ref{thm:pivotal}); exact strong family-wise error
control through a Westfall--Young joint permutation that preserves
cross-coordinate dependence, with a distribution-free Holm fallback
(Theorem~\ref{thm:fwer}); and false discovery rate control under
arbitrary coordinate dependence through Benjamini--Yekutieli and e-BH
(Theorem~\ref{thm:fdr}). Each certified coordinate additionally carries
a location or scale label determined by its dominant evidence channel.
Every reported number derives from executed code and archived outputs;
the experiments cover the selection effect of naive testing at
$\hat\tau$, validity under heavy tails and cross-coordinate dependence,
high-dimensional power, type-label accuracy, and an application to five
financial series surrounding the 2008 crisis.

Section~\ref{sec:background} places ARM among the three lines above;
Section~\ref{sec:method} defines the wrapper; Section~\ref{sec:theory}
states the theory; Sections~\ref{sec:sim}--\ref{sec:real} report
simulations and the financial study; Section~\ref{sec:disc} discusses
limitations.

\section{Background: Attribution Without Certificates}\label{sec:background}

\subsection{Post-detection inference: time axis versus coordinate axis}
\label{sec:timevscoord}

Given detected changepoints, one may ask which are genuine. TUNE
\citep{jia2024tune} and the ART diagnostic \citep[Thm.~5]{cui2026art}
control a family-wise error rate over detected times, algorithm-agnostically
and distribution-freely; selective and sequential procedures
\citep{jewell2022post,saha2026post,saha2026distributionfree} give
post-detection confidence statements about the location. These are guarantees
along the time axis. ARM's Theorem~\ref{thm:fwer} transposes them to
the coordinate axis, applying the same pivotal distribution-free
mechanism across the $d$ coordinates of a single window rather than
across candidate times; the proof pattern is inherited directly from
this literature.

\subsection{Attribution: granularity and guarantee}\label{sec:granularity}

Table~\ref{tab:landscape} places the coordinate-attribution methods on
four axes. \citet{ouerfelli2026posthoc} give the first formal
post-detection attribution: for two predefined coordinate blocks, kernel
maximum-mean-discrepancy tests across $\hat\tau$, calibrated by a
concentration threshold and a holdout split, with
$\mathrm{FWER}\le\alpha_0+L\alpha_1$ by a union bound over $L$ blocks.
Remark~9 of that paper recommends Bonferroni or Holm corrections when
more than two blocks are tested, and its Remark~8 places purely
inter-block dependence changes with invariant marginals outside the
scope of marginal procedures. WAVE \citep{lan2026wave} and related
high-dimensional identification methods
\citep{liu2026simultaneous,zhang2026spectral} attribute at the
coordinate level but through weights or effect rankings, without a
finite-sample false-positive guarantee. ARM advances all four axes at
once: coordinate granularity, exact rather than additively bounded FWER,
rank-permutation rather than concentration-threshold calibration, and FDR
under arbitrary coordinate dependence. The two-block design of
\citet{ouerfelli2026posthoc} is recovered as the aggregation special
case in which a block is scored by the maximum of its coordinate
statistics.

\begin{table}[t]
\centering
\caption{Coordinate-attribution methods on four axes. ARM occupies the
last row.}
\label{tab:landscape}
\small
\resizebox{\textwidth}{!}{%
\begin{tabular}{lcccc}
\toprule
Method & Granularity & Guarantee & Calibration & Dependence \\
\midrule
Ouerfelli et al. & predefined blocks & FWER $\le\alpha_0+L\alpha_1$ & kernel MMD + holdout & independent rows \\
WAVE & coordinate & none (ranking) & block bootstrap & block \\
ART Thm.~5 / TUNE & time (alarms) & exact FWER & rank permutation & independent \\
\textbf{ARM (this paper)} & \textbf{coordinate} & \textbf{exact FWER \& arb.-dep.\ FDR} & \textbf{rank permutation} & \textbf{arbitrary (e-BH)} \\
\bottomrule
\end{tabular}}
\end{table}

\subsection{Multiple testing tools}\label{sec:mttools}

ARM assembles three classical instruments in the rank-permutation
setting: Westfall--Young single-step maxT with the subset-pivotality
condition for strong FWER \citep{westfall1993resampling,meinshausen2008hierarchical},
the exactness of Monte Carlo permutation tests
\citep{hemerik2018exact,lehmann2005testing}, and, for FDR under arbitrary
dependence, Benjamini--Yekutieli \citep{benjamini2001control} and the
recent e-BH procedure \citep{wang2022ebh}.

\section{Methodology: The ARM Wrapper}\label{sec:method}

\subsection{Setting and black-box interface}\label{sec:setting}

The window is $X\in\mathbb R^{n\times d}$: $n$ time points (rows) and $d$
coordinates (columns). Rows are independent; columns may be arbitrarily
dependent. A black-box detector returns an estimated changepoint
$\hat\tau$; ARM never inspects its internals and tolerates an
\emph{arbitrary} $\hat\tau$. For coordinate $j$, the null hypothesis is
\[
H_{0j}:\ X_{1j},\dots,X_{nj}\ \text{are identically distributed (no
marginal change in coordinate } j).
\]
The true-null set is $J_0=\{j: H_{0j}\ \text{holds}\}$; attribution seeks
the complement while controlling false positives over $J_0$.

\subsection{Coordinate rank channels and the max-over-splits statistic}
\label{sec:stat}

Let $R_{\cdot j}$ be the within-column ranks of coordinate $j$ and
$C_{ij}=R_{ij}-(n+1)/2$ the centred ranks. For a split $t$ on a grid
$\mathcal G\subseteq\{m_0,\dots,n-m_0\}$, the location
\citep{wilcoxon1945individual} and scale \citep{mood1954asymptotic}
evidence are
\[
Z^{\mathrm{loc}}_j(t)=\frac{|\sum_{i\le t}C_{ij}|}{\sqrt{t(n-t)(n+1)/12}},
\qquad
Z^{\mathrm{sc}}_j(t)=\frac{|\sum_{i\le t}(C_{ij}^2-\overline{C^2_j})|}
{\sqrt{t(n-t)\,\mathrm{Var}(C^2_j)/n}} .
\]
ARM's coordinate statistic is the maximum over splits and channels,
\begin{equation}\label{eq:M}
\Msf_j=\max_{t\in\mathcal G}\max\{Z^{\mathrm{loc}}_j(t),Z^{\mathrm{sc}}_j(t)\},
\qquad
\Msf^{\mathrm{loc}}_j=\max_t Z^{\mathrm{loc}}_j(t),\quad
\Msf^{\mathrm{sc}}_j=\max_t Z^{\mathrm{sc}}_j(t),
\end{equation}
with $\Msf^{\mathrm{loc}}_j,\Msf^{\mathrm{sc}}_j$ retained for the type
label.

\subsection{Attribution rule and selection-proofness}\label{sec:rule}

ARM attributes coordinate $j$ when $\Msf_j$ exceeds a distribution-free
critical value: $A_\alpha=\{j:\Msf_j>c_{n,\alpha}\}$, where $c_{n,\alpha}$
is the $(1-\alpha)$ quantile of $\Msf$ under a single null coordinate,
computed once by permuting a rank vector (it depends only on $n$ and the
grid, never on the data). The decisive choice is to score by $\Msf_j$ rather
than by $Z_j(\hat\tau)$. Since $Z_j(\hat\tau)\le\Msf_j$ for every
split, no detector output, whether accurate or arbitrarily misplaced,
can raise a true-null coordinate's statistic above the level that the
maximum already attains; attribution is therefore selection-proof by
construction (Theorem~\ref{thm:pivotal}), and $\hat\tau$ affects the
power of the procedure but not its validity. The per-coordinate permutation $p$-value is
$p_j=(1+\#\{\text{null draws}\ge\Msf_j\})/(B+1)$.

\subsection{FWER: Westfall--Young joint permutation}\label{sec:fwer}

A single time permutation $\pi$ is applied to \emph{all} columns at once
(row permutation), preserving cross-coordinate dependence; the reference
is the permutation law of $\max_j\Msf_j(X_\pi)$. The single-step maxT
adjusted $p$-value is
$\tilde p_j=(1+\#\{b:\max_k\Msf_k(X_{\pi_b})\ge\Msf_j\})/(B_{\mathrm{wy}}+1)$,
and ARM rejects $H_{0j}$ when $\tilde p_j\le\alpha$. Under subset
pivotality (Section~\ref{sec:theory}) this controls FWER exactly. A
fully distribution-free fallback applies Holm to the per-coordinate
$p_j$, valid with no dependence assumption at a mild power cost.

\subsection{FDR in high dimensions}\label{sec:fdr}

For large $d$ we control the false discovery rate under \emph{arbitrary}
coordinate dependence. Benjamini--Yekutieli \citep{benjamini2001control}
applied to the permutation $p$-values pays the harmonic penalty
$c_d=\sum_{i\le d}1/i$ and is our robust default. As a modern
alternative we form boosted threshold e-values
$e_j=s^{-1}\mathbf 1\{p_j\le s\}$ (valid since
$\ex_{H_{0j}}e_j=\pr(p_j\le s)/s\le1$) and apply e-BH \citep{wang2022ebh},
which controls FDR under arbitrary dependence \emph{for every choice of}
$s$; the choice affects power only. The mechanics fix the rule: e-BH
rejects $k$ coordinates only if $s^{-1}\ge d/(\alpha k)$, so $s$ must
not exceed $\alpha k/d$ for the smallest number of changed coordinates
one intends to detect, and $p_j\le s$ must be reachable, requiring
$B\ge 1/s$ permutations. In the study below we use $s=0.004$, satisfying this rule for
$k=15$ changed coordinates at $d=200$ ($\alpha k/d=0.0075$) with
$B=1999$; Section~\ref{sec:sim-highdim} reports the sensitivity across
$s$, and we emphasize that FDR control holds for every $s$; the rule
only locates the power boundary.

\subsection{Type label}\label{sec:type}

Each attributed coordinate is labelled \emph{scale} if
$\Msf^{\mathrm{sc}}_j>\Msf^{\mathrm{loc}}_j$ and \emph{location}
otherwise. The label is descriptive and is not part of any error-control
statement; a certified type label would require closed testing, which we
leave to future work.

\begin{figure}[t]
\centering
\resizebox{\textwidth}{!}{%
\begin{tikzpicture}[font=\footnotesize,
  box/.style={draw, rounded corners=2pt, align=center, inner sep=4pt, minimum height=8mm},
  chan/.style={box, fill=blue!8}, frozen/.style={box, fill=orange!12},
  outp/.style={box, fill=green!10}, lab/.style={font=\scriptsize\itshape, text=black!60, align=center},
  arr/.style={-latex, thick, black!70}, node distance=4mm and 8mm]
\node[box] (det) {black-box\\detector $\to\hat\tau$\\(any algorithm)};
\node[box, right=of det] (win) {window $X$\\$n\times d$};
\node[chan, right=of win, yshift=8mm] (loc) {location channel\\$Z^{\mathrm{loc}}_j(t)$};
\node[chan, right=of win, yshift=-8mm] (sc) {scale channel\\$Z^{\mathrm{sc}}_j(t)$};
\node[frozen, right=of loc, yshift=-8mm] (M) {$\Msf_j=\max_t\max(\cdot)$\\(max over splits)};
\node[box, right=of M] (cal) {WY joint perm.\\/ Holm / e-BH};
\node[outp, right=of cal] (out) {certified set $A_\alpha$\\+ type label\\per coordinate};
\node[lab, above=1mm of loc] {ranks only $\Rightarrow$ distribution-free};
\node[lab, below=5mm of sc] {$Z_j(\hat\tau)\le \Msf_j\Rightarrow$ selection-proof (Thm.~1)};
\draw[arr] (det) -- (win); \draw[arr] (win) -- (loc); \draw[arr] (win) -- (sc);
\draw[arr] (loc) -- (M); \draw[arr] (sc) -- (M); \draw[arr] (M) -- (cal);
\draw[arr] (cal) -- (out);
\end{tikzpicture}}
\caption{The ARM wrapper. Any detector supplies $\hat\tau$; ARM scores
each coordinate by a max-over-splits rank statistic (so the certificate
ignores $\hat\tau$) and calibrates by joint permutation (FWER) or e-BH/BY
(FDR), returning a certified coordinate set with type labels.}
\label{fig:pipeline}
\end{figure}
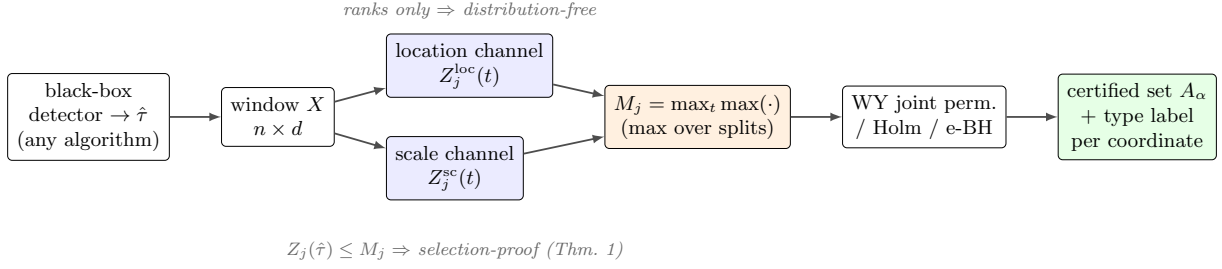

Figure~\ref{fig:pipeline} summarizes the wrapper.

\section{Theory}\label{sec:theory}

\begin{theorem}[Per-coordinate validity, any detector]\label{thm:pivotal}
Under $H_{0j}$, the law of $\Msf_j$ depends only on $(n,\mathcal G)$,
and not on the marginal $F_j$, the other coordinates, or the detector.
Hence for
any $\alpha\in(0,1)$ and \emph{any} rule producing $\hat\tau$,
$\pr(j\in A_\alpha)\le\alpha$.
\end{theorem}

The proof (Appendix~A) is that $\Msf_j$ is a function of coordinate $j$'s
ranks, which under $H_{0j}$ are a uniform random permutation; the
statement holds for every $\hat\tau$ because $\Msf_j$ never uses
$\hat\tau$.

\begin{theorem}[Exact strong FWER]\label{thm:fwer}
Suppose the rows of the true-null sub-array
$(X_{i,J_0})_{i=1}^n$ are independent and identically distributed, so
that the joint law of the null coordinates, including their mutual
dependence, is constant over time. Then
the Westfall--Young single-step maxT of Section~\ref{sec:fwer} controls
the family-wise error rate over $J_0$ in the strong sense,
$\pr(\exists j\in J_0: j\in A_\alpha)\le\alpha$.
\end{theorem}

The condition is subset pivotality: since each $\Msf_j$ depends only on
column $j$, the joint null law of $\{\Msf_j:j\in J_0\}$ under the row
permutation is unaffected by whether the non-null coordinates changed
(Appendix~A). A pure change in the \emph{dependence} among null
coordinates with invariant marginals makes the null rows
non-identically distributed and is out of scope; this is the same
boundary drawn by \citet[Remark~8]{ouerfelli2026posthoc}. When the condition is
in doubt, the Holm fallback on $p_j$ is exact under the marginal nulls
alone, with no assumption on the joint law.

\begin{theorem}[FDR under arbitrary dependence]\label{thm:fdr}
For any joint dependence among the $d$ coordinates, Benjamini--Yekutieli
on $\{p_j\}$ and e-BH on the boosted e-values $\{e_j\}$ each control the
false discovery rate at level $\alpha$.
\end{theorem}

BY controls FDR under arbitrary dependence by the harmonic-penalty
theorem \citep{benjamini2001control}; the $e_j$ are valid e-values, so
e-BH controls FDR with no dependence assumption \citep{wang2022ebh}
(Appendix~A).

\begin{proposition}[Support recovery]\label{prop:power}
Fix a coordinate $j^\star$ with a stochastically ordered marginal change
of fixed size. Then $\Msf_{j^\star}\to\infty$ in probability as
$n\to\infty$, so $\pr(j^\star\in A_\alpha)\to1$; consequently, with a
bounded number of changed coordinates, the FWER and FDR procedures
recover the full changed set with probability tending to one.
\end{proposition}

Proposition~\ref{prop:power} (proof sketch in Appendix~A) states power as
a limit; the finite-sample behaviour is the simulation study, reported in
full.

\section{Simulation Studies}\label{sec:sim}

All numbers below are read from the generated result files
(\texttt{run\_experiments.py}). Windows have $n=120$, changepoint
$\tau=60$, margin $m_0=10$, nominal $\alpha=0.10$; the null reference uses
$B=999$ rank permutations and the Westfall--Young reference
$B_{\mathrm{wy}}=499$. Monte Carlo standard errors are about $0.02$ at
$250$--$300$ replications.

\subsection{Testing at an estimated changepoint: the selection effect}
\label{sec:sim-killshot}

\begin{figure}[t]
\centering
\includegraphics[width=\textwidth]{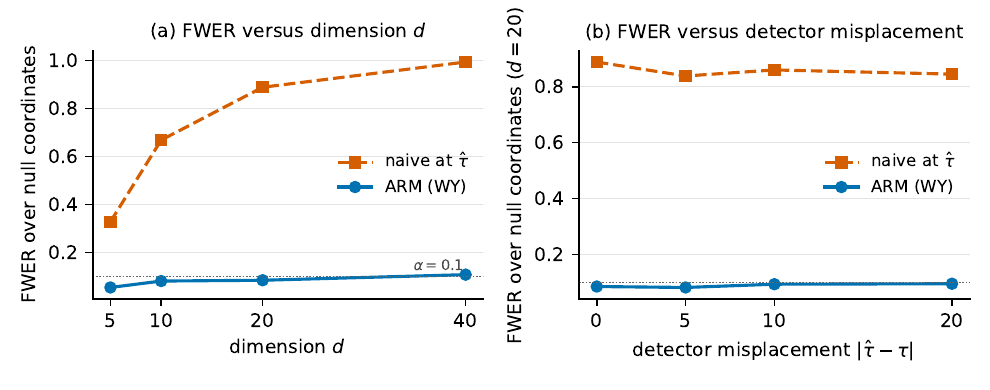}
\caption{The selection effect. (a) FWER over the null coordinates
against dimension, with the detector at $\hat\tau=\tau$: naive testing
inflates toward one while ARM maintains the nominal $0.10$. (b) FWER
against detector misplacement $|\hat\tau-\tau|$ at $d=20$: ARM is flat
because its statistic does not involve $\hat\tau$, whereas the naive
test remains inflated at every misplacement.}
\label{fig:killshot}
\end{figure}

Each window contains two changed coordinates, so that the detector
locks onto a meaningful $\hat\tau$, and the family-wise error is
measured over the remaining $d-2$ true-null coordinates. ARM is
compared against the naive per-coordinate procedure, in which each
coordinate's statistic at the estimated split is referred to a
fixed-split null distribution. Figure~\ref{fig:killshot} and
Table~\ref{tab:killshot} report the results. The naive error reaches
$0.33$ at $d=5$ and increases to $0.67$ at $d=10$ and $0.89$ at
$d=20$: the detector selects the split at which the aggregate evidence
peaks, and that selection contaminates every per-coordinate test
performed there. ARM maintains the nominal $0.10$ throughout, with
entries between $0.052$ and $0.108$ across all sixteen cells, and is
flat in the detector misplacement $|\hat\tau-\tau|$ because $\Msf_j$
does not involve $\hat\tau$, whereas the naive test remains inflated at
every misplacement.

Two remarks concern the fairness of this comparison. First, the naive
procedure is not an artificial comparator: treating $\hat\tau$ as known
and applying a per-coordinate two-sample test at it is precisely the
documented double-dipping practice that motivates post-detection
inference \citep{jewell2022post,jia2024tune}. Second, the principled
correction of that practice, namely an adjustment for the maximization
over splits, is exactly what ARM implements, not approximately through
a Bonferroni penalty over the grid but exactly through the permutation
law of the maximum. The inflation measured here and the validity of
the method are consequences of the same inequality
$Z_j(\hat\tau)\le\Msf_j$.

\begin{table}[t]
\centering
\caption{Selection effect: FWER over the null coordinates, $600$
replications ($t_3$ marginals, two changed coordinates, aggregate-CUSUM
detector). Naive per-coordinate testing at $\hat\tau$ inflates with
$d$; ARM maintains the nominal level. Columns are detector
misplacements $|\hat\tau-\tau|$.}
\label{tab:killshot}
\small
\begin{tabular}{l cccc c cccc}
\toprule
& \multicolumn{4}{c}{naive at $\hat\tau$} & & \multicolumn{4}{c}{ARM (WY)} \\
\cmidrule(r){2-5}\cmidrule(l){7-10}
$d$ & 0 & 5 & 10 & 20 & & 0 & 5 & 10 & 20 \\
\midrule
5  & 0.33 & 0.27 & 0.31 & 0.29 & & 0.055 & 0.060 & 0.083 & 0.052 \\
10 & 0.67 & 0.62 & 0.56 & 0.56 & & 0.082 & 0.097 & 0.072 & 0.077 \\
20 & 0.89 & 0.84 & 0.86 & 0.84 & & 0.085 & 0.082 & 0.093 & 0.095 \\
40 & 0.99 & 0.98 & 0.99 & 0.98 & & 0.108 & 0.103 & 0.100 & 0.107 \\
\bottomrule
\end{tabular}
\end{table}

\subsection{Validity under heavy tails, dependence, and any detector}
\label{sec:sim-validity}

Table~\ref{tab:validity} sweeps $d=10$ with three changed coordinates
over four marginals, including the heavy-tailed and skewed regimes that
motivate distribution-free tools \citep{tang2026online,li2026robust},
three cross-coordinate dependence structures, and three detectors: the
true $\tau$, an aggregate-CUSUM estimate, and a deliberately misplaced
$\hat\tau=\tau+10$. Every WY and Holm FWER entry
sits at or within Monte Carlo error of $0.10$ (range $0.032$--$0.108$
over all $36$ cells), and every BY FDR entry is far below (range
$0.010$--$0.029$, reflecting three true signals among ten coordinates).
The detector variant changes the entries by at most Monte Carlo error,
confirming that the guarantee does not depend on how $\hat\tau$ was
produced, as Theorem~\ref{thm:pivotal} predicts.
Table~\ref{tab:validity} reports the CUSUM-detector column; the
true-$\tau$ and misplaced columns differ by at most $0.03$ and are
contained in the archived results.

\begin{table}[t]
\centering
\caption{Validity at $d=10$, three changed coordinates, CUSUM detector,
$250$ replications (MC-SE $\approx0.019$ for FWER). WY and Holm control
FWER; BY controls FDR. Heavy tails and cross-coordinate dependence do not
break control.}
\label{tab:validity}
\small
\begin{tabular}{ll ccc}
\toprule
Marginal & Dependence & WY FWER & Holm FWER & BY FDR \\
\midrule
Gaussian  & independent & 0.064 & 0.072 & 0.018 \\
          & equicorrelated & 0.076 & 0.072 & 0.018 \\
          & factor & 0.088 & 0.092 & 0.022 \\
$t_3$     & independent & 0.080 & 0.084 & 0.021 \\
          & equicorrelated & 0.076 & 0.088 & 0.023 \\
          & factor & 0.072 & 0.076 & 0.020 \\
Cauchy    & independent & 0.044 & 0.036 & 0.011 \\
          & equicorrelated & 0.060 & 0.060 & 0.016 \\
          & factor & 0.064 & 0.076 & 0.019 \\
Lognormal & independent & 0.072 & 0.084 & 0.022 \\
          & equicorrelated & 0.044 & 0.052 & 0.013 \\
          & factor & 0.072 & 0.084 & 0.021 \\
\bottomrule
\end{tabular}
\end{table}

\subsection{Power and granularity}\label{sec:sim-power}

We compare ARM's certified coordinate attribution against two baselines
on $d=20$ with three changed coordinates: a simplified two-block test
in the style of \citet{ouerfelli2026posthoc}, implemented with the
energy distance \citep{szekely2013energy} in place of the kernel
maximum mean discrepancy and without the holdout split, which certifies
only a coarse block; and a WAVE-style fusion \citep{lan2026wave} that
returns a coordinate ranking without error control, scored by the AUC
of signal coordinates over null coordinates. Table~\ref{tab:power}
reports a strong ($\delta=1.8$) and a weak ($\delta=1.1$) signal
regime. Two conclusions follow. First, ARM's certified attribution is
essentially exact under the strong signal, with F1 between $0.985$ and
$0.987$, and degrades gracefully under the weak signal, with F1 between
$0.88$ and $0.99$. Second, the comparison concerns the form of the
output rather than detection quality: at these signal levels all three
methods locate the changed coordinates, and WAVE's ranking AUC remains
above $0.99$, but WAVE returns a ranking without a certificate, the
block test certifies only a coarse block, and only ARM returns
individual coordinates with a finite-sample guarantee.

\begin{table}[t]
\centering
\caption{Attribution quality at $d=20$, three changed coordinates, $200$
replications. ARM's coordinate-level F1 versus a block test's coarse
block-hit rate and a WAVE-style ranking AUC. The baselines find the
signals; only ARM certifies individual coordinates.}
\label{tab:power}
\small
\begin{tabular}{ll ccc}
\toprule
Signal & Marginal & ARM coord.\ F1 & block hit (coarse) & WAVE rank AUC \\
\midrule
\multirow{3}{*}{$\delta=1.8$}
 & Gaussian  & 0.987 & 1.000 & 1.000 \\
 & $t_3$     & 0.985 & 1.000 & 1.000 \\
 & Lognormal & 0.986 & 0.995 & 1.000 \\
\addlinespace
\multirow{3}{*}{$\delta=1.1$}
 & Gaussian  & 0.980 & 1.000 & 1.000 \\
 & $t_3$     & 0.877 & 0.945 & 0.995 \\
 & Lognormal & 0.986 & 0.890 & 1.000 \\
\bottomrule
\end{tabular}
\end{table}

\subsection{High-dimensional FDR}\label{sec:sim-highdim}

\begin{figure}[t]
\centering
\includegraphics[width=0.72\textwidth]{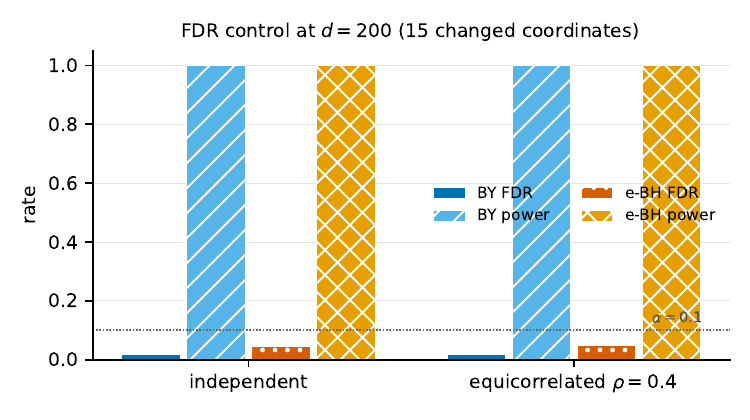}
\caption{High-dimensional attribution ($d=200$, $15$ changed). BY and
e-BH both hold FDR at or below the nominal $\alpha=0.10$ under
independent and equicorrelated coordinates while recovering the full
support (power $1.00$).}
\label{fig:highdim}
\end{figure}

At $d=200$ with $15$ changed coordinates (Figure~\ref{fig:highdim}),
both FDR procedures control the rate under independent and
equicorrelated ($\rho=0.4$) coordinates while recovering the full
support. BY reports FDR $0.019$ (independent) and $0.017$
(equicorrelated) at power $1.00$; e-BH, with the boosted threshold
$s=0.004$ and $B=1999$, reports FDR $0.043$ and $0.048$ at power $1.00$.
BY is the tighter default here; e-BH is the option when only an e-value,
not a $p$-value, is available or when combining evidence across
analyses. Its power depends on $s$ relative to the sparsity exactly as
the rule of Section~\ref{sec:fdr} predicts. In an archived sensitivity
run under the equicorrelated design, the values
$s\in\{0.002,0.004,0.006\}$, all below the boundary
$\alpha k/d=0.0075$, give power $1.00$ at FDR between $0.027$ and
$0.060$, while $s=0.008$, just above the boundary, reduces power to
$0.66$ at FDR $0.082$. The FDR remains controlled at every $s$, as
Theorem~\ref{thm:fdr} guarantees; only power is affected.

\subsection{Type labels and ablations}\label{sec:sim-types}

Type labelling is nearly perfect: over $300$ replications with four
location-shift and four scale-shift coordinates, attributed
location coordinates were labelled \emph{location} $1200/1200$ times and
attributed scale coordinates \emph{scale} $1058/1060$ times (two
mislabels). Ablations confirm robustness: FWER and power are unchanged
across $B\in\{199,499,999\}$ null permutations (FWER $0.053$, power
$1.00$ throughout), and coarsening the split grid from step $1$ to step
$4$ leaves FWER in $[0.073,0.087]$ at power $1.00$, cutting the
permutation cost roughly fourfold.

\section{Real Data: Attributing the 2008 Regime Shift}\label{sec:real}

We take five daily financial series from FRED \citep{fred2026}, namely the
NASDAQ Composite, the 10-year Treasury yield, WTI crude oil, the
USD/EUR rate, and the VIX, taken as log-return or first-difference
coordinates over June 2007 to June 2009, and analyse a $120$-day window
centred on the Lehman Brothers collapse of 15 September 2008. To make the
family-wise guarantee demonstrable we append three \emph{control}
coordinates that carry no regime change: an i.i.d.\ resample of the
pre-window exchange-rate returns and two i.i.d.\ noise series matched in
scale. The injection serves a methodological purpose. The real data provide
no ground-truth null coordinates, since all five asset classes
plausibly changed in 2008, and without controls the error-control claim
would be unfalsifiable on this window; the controls, built from the
window's own quiet period and fully disclosed, supply true-null
coordinates on which a false attribution could occur. A black-box variance-CUSUM detector locates
$\hat\tau$ two days after Lehman; the score reference and thresholds are
those of Section~\ref{sec:sim}, with nothing tuned on the financial data.

\begin{figure}[t]
\centering
\includegraphics[width=\textwidth]{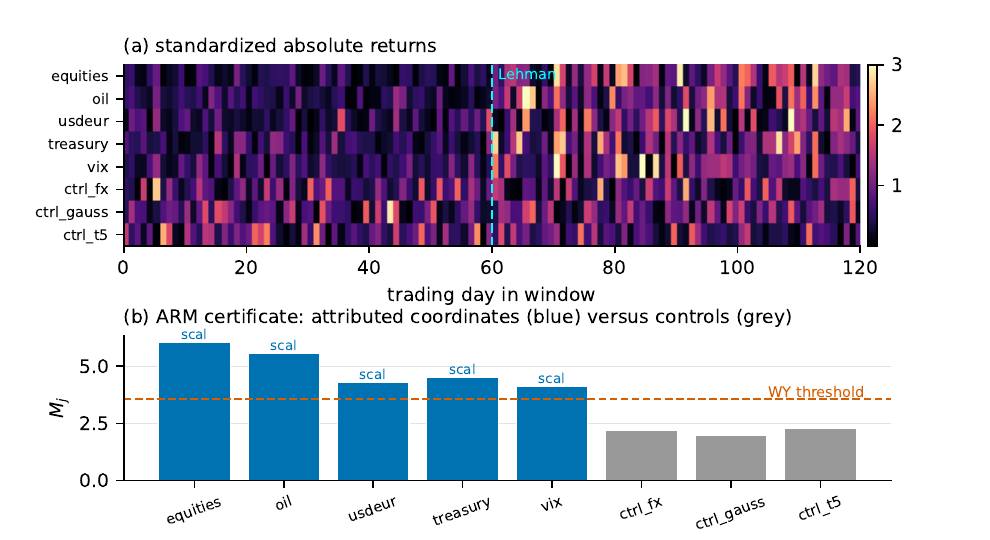}
\caption{Attributing the 2008 regime shift. (a) Standardized absolute
returns: a volatility jump appears across the five financial coordinates
after Lehman (dashed line) while the three controls stay flat. (b) ARM
certificate: the five asset classes (blue) clear the WY threshold
(dashed) with scale-type labels; the three controls (grey) do not.}
\label{fig:panel}
\end{figure}

ARM attributes a change to all five financial coordinates and to none of
the three controls, under WY, Holm, and BY alike
(Table~\ref{tab:panel}, Figure~\ref{fig:panel}). Every attributed
coordinate is labelled \emph{scale}: the crisis is read as a volatility
regime shift. Mean effects are present, notably the fall in equities,
but the scale channel dominates the location channel in every
coordinate, which is the descriptive content of the label. The evidence
orders the asset classes by the size of the volatility jump (equities
$\Msf=6.0$, oil $5.6$, Treasury $4.5$, USD/EUR $4.3$, VIX $4.1$), all
clearing the WY threshold $c=3.6$, while the three controls
($\Msf=2.0$--$2.3$) fall well short. The detector's two-day mislocation
perturbs none of these conclusions, because the certificate is computed
from $\Msf_j$ rather than from the statistic at $\hat\tau$.

\begin{table}[t]
\centering
\caption{ARM on the 2008 window: coordinate statistic $\Msf_j$, type
label, and attribution decision (WY, $\alpha=0.10$, threshold $c=3.6$).
Five asset classes are certified as scale changes; three injected
controls are correctly excluded.}
\label{tab:panel}
\small
\begin{tabular}{lccc}
\toprule
Coordinate & $\Msf_j$ & type & attributed? \\
\midrule
equities (NASDAQ) & 6.04 & scale & yes \\
oil (WTI) & 5.56 & scale & yes \\
Treasury (10y) & 4.50 & scale & yes \\
USD/EUR & 4.29 & scale & yes \\
VIX & 4.11 & scale & yes \\
\addlinespace
control (fx resample) & 2.17 & --- & no \\
control (Gaussian) & 1.99 & --- & no \\
control ($t_5$) & 2.26 & --- & no \\
\bottomrule
\end{tabular}
\end{table}

\section{Discussion}\label{sec:disc}

Four limitations bound the claims. First, exact error control assumes
independent rows; under serial dependence a block-permutation variant
restores validity only approximately, and a rank theory for dependent
rows \citep{banerjee2026markov} is the natural next step. Second, the
type label is descriptive: a certified label would require closed
testing across the location and scale channels, which we leave open.
Third, a pure change in the dependence among coordinates with invariant
marginals lies outside the marginal-attribution model, exactly as in
\citet[Remark~8]{ouerfelli2026posthoc}, and its detection requires a
dedicated dependence test. Fourth, the predefined-block setting of
\citet{ouerfelli2026posthoc} constitutes structural prior information
that ARM does not exploit; when blocks are genuinely known, aggregating
ARM statistics within blocks recovers that setting as a special case.

Two directions follow. ARM and a distribution-free localization set
\citep[e.g.][]{pcc_arc} compose into a complete post-detection toolkit
that reports when the change occurred, with a confidence set, and which
coordinates changed, with a certificate; the rank channels are shared
between the two procedures, inherited from the ART-NN line
\citep{pcc_artnn}. An online variant would replace the permutation
reference with an anytime-valid one \citep{lee2026fdr}. Finally, an
estimation layer such as WAVE \citep{lan2026wave}, which produces
coordinate weights without a guarantee, pairs naturally with ARM as its
inference layer: the estimation layer nominates candidate coordinates
and the inference layer certifies them.

\section{Conclusion}\label{sec:conc}

Detection establishes when a change occurred; ARM establishes which
coordinates changed, with certificates that an incorrectly estimated
changepoint cannot invalidate. Scoring each coordinate by a
max-over-splits rank statistic renders attribution selection-proof by
construction; a joint permutation delivers exact family-wise error
control while preserving cross-coordinate dependence, and
Benjamini--Yekutieli and e-BH extend the guarantee to false discovery
rate control under arbitrary dependence in high dimensions. The
certificate is valid under heavy tails and under any detector, and on
the 2008 data it refines a market-wide alarm into the statement that
every asset class underwent a volatility change while the injected
controls did not.

\section*{Declarations}

\paragraph{Data availability.} All results are reproducible from the
accompanying code (\texttt{arm\_core.py}, \texttt{run\_experiments.py},
\texttt{real\_data\_panel.py}; fixed seeds). The five financial series
are public FRED series (NASDAQCOM, DGS10, DCOILWTICO, DEXUSEU, VIXCLS)
retrieved from \url{https://fred.stlouisfed.org/}.


\paragraph{Conflict of interest.} The authors declare no conflict of
interest.

\appendix
\section{Proofs}\label{app:proofs}

\subsection{Proof of Theorem~\ref{thm:pivotal}}
Fix $j$ and condition on $H_{0j}$: the entries $X_{1j},\dots,X_{nj}$ are
i.i.d., so their rank vector $R_{\cdot j}$ is uniform on the permutations
of $\{1,\dots,n\}$, independent of the coordinate's marginal $F_j$ and of
every other column. The statistic $\Msf_j$ in \eqref{eq:M} is a fixed
function of $R_{\cdot j}$ alone, so its law depends only on $n$ and the
grid $\mathcal G$; the critical value $c_{n,\alpha}$ is the exact
$(1-\alpha)$ quantile of that law, whence $\pr(\Msf_j>c_{n,\alpha})\le
\alpha$. Because $\Msf_j$ does not involve $\hat\tau$ at any point, the
bound holds simultaneously for every rule that could have produced
$\hat\tau$; in particular $Z_j(\hat\tau)\le\Msf_j$ shows that testing at
the estimated split can only be less significant than the max, never
more. \qed

\subsection{Proof of Theorem~\ref{thm:fwer}}
Write $J_0$ for the true-null coordinates and let $\pi$ be a uniform
random time permutation applied jointly to all columns. Under the stated
condition the sub-array $(X_{i,J_0})_{i=1}^n$ is exchangeable in $i$, so
for the $J_0$-columns the row-permuted window $X_\pi$ has the same joint
law as $X$; consequently the joint law of $\{\Msf_j(X_\pi):j\in J_0\}$
equals that of $\{\Msf_j(X):j\in J_0\}$, and neither depends on the
values of the non-null columns because each $\Msf_j$ is a function of
column $j$ only. This is the subset-pivotality condition of
\citet{westfall1993resampling}. The single-step maxT reference
$\max_k\Msf_k(X_\pi)$ dominates $\max_{j\in J_0}\Msf_j(X_\pi)$, so the
adjusted $p$-values $\tilde p_j$ are jointly valid and
$\pr(\exists j\in J_0:\tilde p_j\le\alpha)\le\alpha$ by the standard
permutation argument \citep{hemerik2018exact,meinshausen2008hierarchical}.
\qed

\subsection{Proof of Theorem~\ref{thm:fdr}}
The permutation $p$-values $p_j$ are each super-uniform under $H_{0j}$
(Theorem~\ref{thm:pivotal}), so Benjamini--Yekutieli with the harmonic
factor $c_d=\sum_{i\le d}1/i$ controls the FDR at $\alpha$ for arbitrary
joint dependence \citep{benjamini2001control}. For the e-BH branch,
$e_j=s^{-1}\mathbf 1\{p_j\le s\}$ satisfies
$\ex_{H_{0j}}[e_j]=\pr(p_j\le s)/s\le 1$ since $p_j$ is super-uniform, so
each $e_j$ is a valid e-value and e-BH controls the FDR at $\alpha$ under
arbitrary dependence \citep{wang2022ebh}. \qed

\subsection{Proof sketch of Proposition~\ref{prop:power}}
For a coordinate with $\rho=\pr(X'>X)\neq\tfrac12$ across a fixed-fraction
split, the Wilcoxon evidence at the true split is a two-sample
$U$-statistic with mean of order $\sqrt n\,|\rho-\tfrac12|$, so
$\Msf_{j^\star}\ge Z^{\mathrm{loc}}_{j^\star}(\tau)\to\infty$ in
probability while the null critical value $c_{n,\alpha}$ grows only as
$\sqrt{\log n}$ (the maximum of $O(n)$ standardized rank CUSUMs with
sub-Gaussian tails), and the Westfall--Young critical value adds only a
$\sqrt{\log d}$-order term with $d$ held fixed. Hence
$\pr(j^\star\in A_\alpha)\to1$, and with $d$ fixed and a bounded number
of changed coordinates a union bound gives simultaneous recovery. The scale channel is analogous with squared centred ranks.
Constants are not optimized; the finite-sample behaviour is
Section~\ref{sec:sim}. \qed

\bibliographystyle{plainnat}
\bibliography{arm}

\end{document}